\documentclass[10pt]{iopart} \usepackage{graphicx,epsfig}

\begin{document}
\title[Dark soliton in a quasi-1D BEC perturbed by an optical lattice]
{Dynamical instability of a dark soliton in a
quasi-one-dimensional Bose-Einstein condensate perturbed by an
optical lattice}

\author{N.G. Parker$^{\dag}$, N.P. Proukakis$^{\dag}$, C.F Barenghi$^{\ddag}$, and C.S. Adams$^{\dag}$}
\address{$^\dag$ Department of Physics, University of Durham, South
Road, Durham DH1 3LE, United Kingdom}
\address{$^\ddag$ School of Mathematics and Statistics, University of Newcastle, Newcastle upon Tyne NE1 7RU, United
Kingdom}

\ead{n.g.parker@durham.ac.uk}
\begin{abstract}
The motion of a dark soliton is investigated in a one-dimensional
dilute Bose-Einstein condensate confined in a harmonic trap and an
optical lattice.  The harmonic trap induces a dynamical
instability of the soliton, culminating in sound emission.  The
presence of the optical lattice enhances the instability, and in
addition, dephases the emitted sound waves, thus preventing
stabilisation of the soliton by sound reabsorption.  This
instability can be probed experimentally by monitoring the soliton
oscillations under various lattice configurations, which can be
realised by changing the intensity and angle between the laser
beams that form the lattice. For short enough times, such that the
emitted sound does not reinteract with the soliton, the power
emitted by the soliton is found to be proportional to the square
of the local soliton acceleration, which is in turn proportional
to the deformation of the soliton profile.
 \end{abstract}
\maketitle

\section{Introduction}
The realisation of dilute atomic Bose-Einstein condensates (BEC)
in multi-dimensional optical lattices has opened the door to study
novel quantum effects within a pure and highly tunable system.
These include the quantum phase transition from superfluid state
to Mott insulator \cite{greiner}, interference of coherent matter
waves \cite{anderson1}, Bloch oscillations \cite{morsch}, number
squeezed states \cite{orzel}, and arrays of Josephson junctions
\cite{cataliotti}. Following the recent experimental creation of
dark solitary waves in harmonically trapped BEC
\cite{burger,denschlag,dutton} it is a now a realistic prospect to
consider these macroscopic excitations in an optical lattice
geometry.

A dark soliton is a one-dimensional localised wavepacket
consisting of a notch in the ambient density and a phase slip
across its centre.  It is supported in repulsive/defocussing
nonlinear media where the shape of the wavepacket is preserved by
a balance between dispersive effects and nonlinearity.  Dark
solitons have been investigated thoroughly, both experimentally
and theoretically, within the analogous field of nonlinear optics
\cite{kivshar1}. However, the nature of trapped BEC poses new
questions for the dynamics and stability of dark solitons. Thermal
instabilities, arising from interaction with the thermal cloud,
dissipate energy from the soliton, causing it to accelerate to the
edge of the condensate and ultimately disappear \cite{fedichev},
while slow solitons are sensitive to quantum fluctuations
\cite{dziarmaga}.

A dark soliton experiences two kinds of {\it dynamical}
instabilites. Firstly, a dark soliton, which is strictly a
one-dimensional object,  is prone to transverse instabilities
\cite{feder,brand,muryshev1} when embedded in a two or
three-dimensional system.  For example, dark solitary waves under
weak transverse confinement have been observed to bend via the
``snake" instability and ultimately decay into vortices, in both
BECs \cite{anderson2} and nonlinear optics \cite{mamaev}. However,
by employing tight transverse confinement to freeze out the
transverse degrees of freedom,
 quasi-one-dimensional condensates can be created
\cite{gorlitz,schreck,ott,hansel}, in which dark solitons are
expected to have greatly enhanced lifetimes \cite{muryshev2}. The
second dynamical instability arises from the fact that a dark
soliton is a solution of a homogeneous system.  The longitudinal
confinement featured in BEC experiments breaks the integrability
of the system, rendering the soliton unstable to motion through
the inhomogeneous background density, and leading to decay via
sound emission \cite{busch,huang}. In previous work we studied
this decay mechanism within a harmonic geometry, and isolated two
limiting cases.  If the emitted sound can escape the system, the
power radiated by the dark soliton is proportional to the local
soliton acceleration squared \cite{parker1}, which in turn is
intimately linked to the deformation of the soliton profile caused
by the emitted sound waves \cite{parker2}. In the opposite case of
an infinite harmonic system, sound reabsorption occurs and the
soliton decay becomes fully stabilised. To control this
experimentally, we suggested creating a soliton in a tight dimple
trap within a weaker harmonic trap. In such a geometry, the depth
of the dimple (relative to the chemical potential) provides a
sensitive handle on the soliton-sound interactions, which can be
observed experimentally by monitoring changes in the soliton
amplitude and period \cite{parker1}.

Recent work by Frantzeskakis {\it et al.}
\cite{kevrekidis,theocharis} has discussed the dynamics of
slowly-moving dark solitons within optical lattices, including the
effects of sound emission. The entire range of lattice
periodicities were considered in \cite{theocharis}, and, where
appropriate, analytic models for the soliton dynamics were
derived. In our work we perform detailed quantitative measurements
of the sound emission from the soliton and show that the
acceleration squared law discussed above is valid (prior to
reabsorption of sound). We concentrate on much faster solitons
than \cite{theocharis}, and discuss in detail a range of lattice
periodicities that can be readily achieved experimentally, where
the lattice wavelength is comparable to, or slightly larger than,
the soliton width. Given the number of current BEC experiments
featuring optical lattices
\cite{greiner,anderson1,morsch,orzel,cataliotti,denschlag_lattice}
and the rapid soliton decay that can be induced in certain limits,
this system may be preferential to investigate soliton-sound
interactions than the dimple trap proposed in \cite{parker1}.

In Section 2 we introduce the system under consideration, the
basic properties of a dark soliton, and the Gross-Pitaevskii
equation used to simulate the soliton dynamics. In Section 3 we
illustrate the dynamical behaviour of a dark soliton in an optical
lattice, in particular the emission of sound, and compare this to
the behaviour in the absence of the optical lattice. The
dependence of the soliton dynamics on the lattice height and
periodicity is discussed in Sections 4 and 5, and in Section 6 we
analyse in detail the soliton dynamics and quantify the power
radiated. Finally, in Section 7, we make some concluding remarks.

\section{Theory}

The dynamics of a trapped atomic BEC at low temperatures can be
well described by the mean-field Gross-Pitaevskii (GP) equation.
Assuming a quasi-one-dimensional condensate (oriented along the
{\it x}-axis), the GP equation reduces to its one-dimensional
form,
\begin{eqnarray}
i\hbar \frac{\partial \psi(x,t)}{\partial
t}=\left(-\frac{\hbar^2}{2m}\frac{\partial^2}{\partial
x^2}+V(x)+g|\psi(x,t)|^2-\mu\right)\psi(x,t).
\end{eqnarray}
where $\psi(x)$ is the mean-field order parameter, $m$ is the
atomic mass, and $\mu=ng$ is the chemical potential of the
effectively one-dimensional system. The nonlinear coefficient $g$
arises from the interatomic interactions and has an effective
one-dimensional form, $g=2\hbar^2a/(ml_\perp ^2)$, where
$\l_\perp=\sqrt{\hbar/(m\omega_\perp)}$ is the transverse harmonic
oscillator length and $a$ is the three-dimensional {\it s}-wave
scattering length. Repulsive interactions, $g>0$, are generally
necessary to support a dark soliton, and will be assumed from now
on. However, dark soliton-like structures can, under appropriate
conditions, be supported by attractive interactions
\cite{kevrekidis2}.

We consider an external potential of the form,
\begin{eqnarray}
V(x)=\frac{1}{2}m\omega_x^2x^2+V_0 \sin^2(2\pi x/\lambda).
\end{eqnarray}
The first term represents a harmonic magnetic trap with frequency
$\omega_x$ and the second term denotes an optical lattice with
height $V_0$ and periodicity $\lambda/2$, where $\lambda$ is
determined by the wavelength and relative angle between the laser
beams forming the lattice. Figure 1 shows a schematic of the trap
configuration and initial condensate density.

\begin{figure}
\begin{center}
\includegraphics[width=7cm]{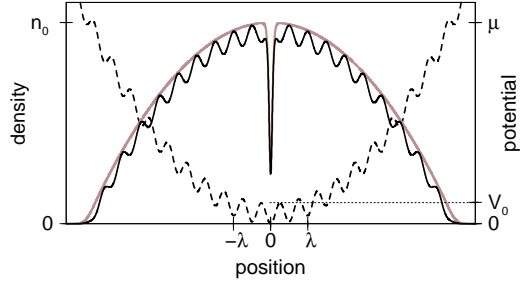}
\end{center}
\caption{Schematic of the initial condensate density (solid bold
line), in relation to the peak condensate density $n_0$, under the
confinement of a harmonic trap and optical lattice (dashed line),
with a non-stationary soliton located at the origin. The optical
lattice has height $V_0$ and wavelength $\lambda$. The
corresponding density in the absence of the optical lattice is
also illustrated (solid grey line).}
\end{figure}

On a uniform background $n_0$, the wavefunction of a dark soliton
with speed $v$ and position $(x-vt)$ is given by,
\begin{eqnarray}
\psi(x,t)=\sqrt{n_0}e^{-i(\mu/\hbar)t} \left( \beta \tanh \left[
\beta
\frac{\left(x-vt\right)}{\xi}\right]+i\left(\frac{v}{c}\right)\right).
\end{eqnarray}
where $\beta=\sqrt{1-(v/c)^2}$ and $\xi=\hbar/\sqrt{\mu m}$ is the
condensate healing length corresponding roughly to the size of the
soliton \cite{jackson}. The soliton speed $v$ depends on its depth
$n_d$ relative to the background density and the phase slip $S$
across its centre via $v=\sqrt{n_0-n_d}=c\cos (S/2)$, with the
maximum speed set by the Bogoliubov speed of sound
$c=\sqrt{\mu/m}$.  The renormalised soliton energy (after
subtracting off the background fluid contribution) is given by
\cite{kivshar1},
\begin{eqnarray}
E_s=\frac{4}{3}\hbar
n_0^{3/2}\left[1-\left(\frac{v}{c}\right)^2\right]^{3/2}.
\end{eqnarray}
A stationary dark soliton reaches zero density ($n_d=n_0$),
features an abrupt $\pi$ phase slip, and maximum energy, whereas a
soliton with maximum speed $c$ has no phase or density contrast,
and therefore zero energy. Note that, to first order, a dark
soliton behaves dynamically like a classical particle
\cite{morgan,reinhardt}.

Throughout this paper we employ harmonic oscillator units, where
length is measured in terms of the harmonic oscillator length
$l_x=\sqrt{\hbar/(m\omega_x)}$ and time in units of
$\omega_x^{-1}$.  In addition a fixed harmonic trap is always
considered, corresponding to $\mu=70\hbar\omega_x$, and
simulations are performed for fixed chemical potential. We will
assume realistic experimental parameters: $\omega_x=2\pi\times5
{\rm Hz}$, $\omega_\perp=250\omega_x$, and
$\mu_{3D}=8\hbar\overline{\omega}$, where
$\overline{\omega}=(\omega_x \omega_\perp^2)^{1/3}$.  Then, for a
$^{87}{\rm Rb}$ ($^{23}{\rm Na}$) condensate, our space and time
units correspond to $l_x=4.8~(9.4)~{\rm \mu m}$ and
$\omega_x^{-1}=0.03{\rm ms}$. Furthermore these parameters roughly
relate to a healing length $\xi=0.58~(1.13)~{\rm \mu m}$, speed of
sound $c=2.5~(1.3)~{\rm mm/ s}$, one-dimensional peak condensate
density $n_{1D}=2 \times 10^7~(5 \times 10^7)~{\rm m^{-1}}$, and
total number of atoms $N=4000 ~(18000)$. All effects considered in
this paper manifest themselves on timescales smaller than the $1$s
anticipated soliton thermodynamic lifetime \cite{fedichev}.

In order to realistically probe a range of optical lattice
periodicities, we envisage employing two almost co-propagating
laser beams, of wavelength
 $\lambda_{\rm laser}$. By varying the angle between the beams
 $\theta$,
 the effective lattice wavelength can be increased beyond the actual laser wavelength, $\lambda \rightarrow
\lambda_{\rm laser}/\sin\theta$. Assuming a laser wavelength
$\lambda_{\rm laser} \sim 0.8 {\rm \mu m}$ (typically produced by
a diode laser), then lattice wavelengths up to the order of
$\lambda \sim 20 {\rm \mu m}$ can easily be produced,
corresponding to
 $\lambda \sim 5 {\it l_x}$ for $^{87} {\rm Rb}$.

\section{Soliton dynamics in an optical lattice and sound emission}
\begin{figure}
\begin{center}
\includegraphics[width=9cm]{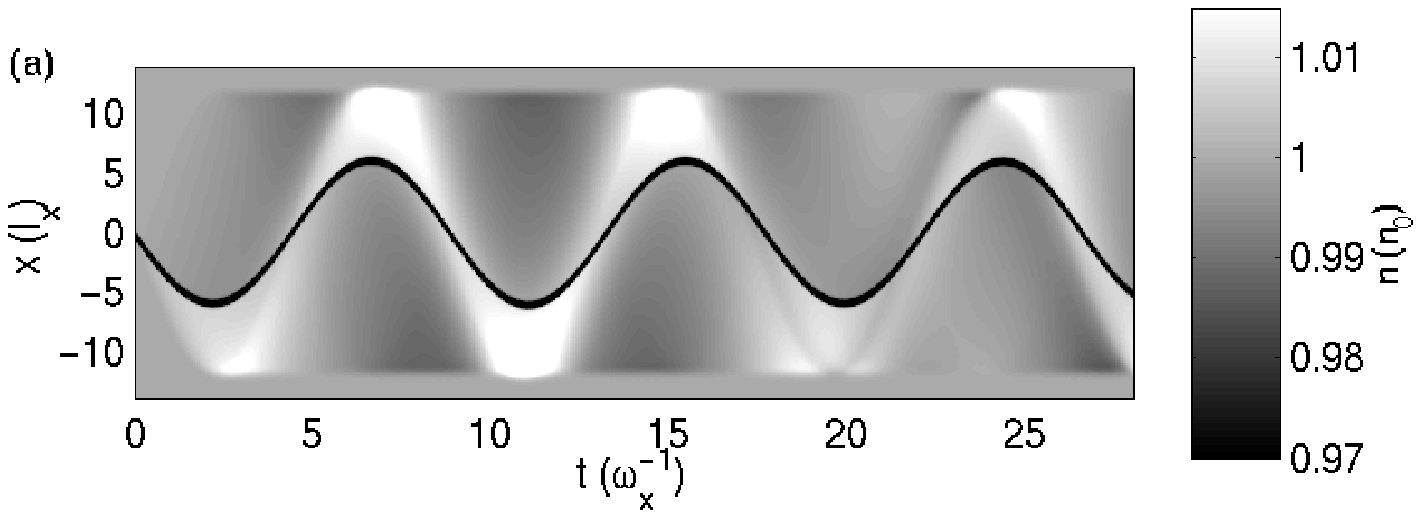}
\includegraphics[width=9cm]{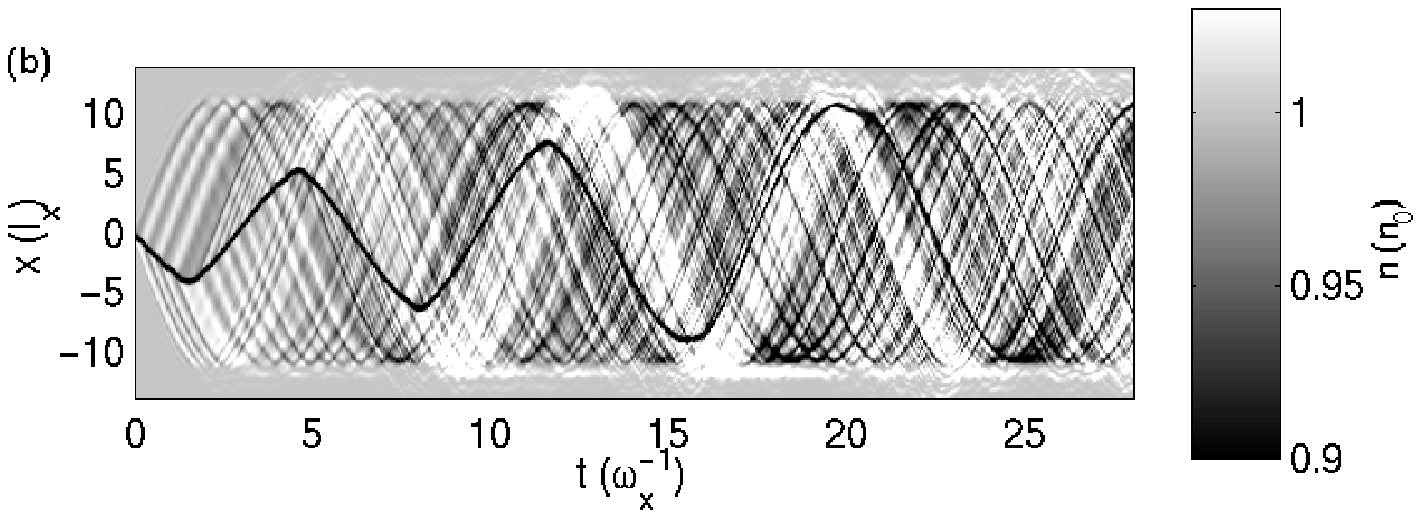}
\end{center}
\caption{Space-time carpet plots of the renormalised condensate
density (actual density minus background density) for (a) harmonic
confinement only and (b) the same harmonic trap plus an optical
lattice with $V_0=0.2\mu$ and $\lambda=2.4 l_x$. Both cases start
with a dark soliton at the origin with speed $v=0.5c$. Sound waves
are clearly emitted by the dynamically unstable soliton, not only
as it climbs the trap walls but also as it traverses the lattice
sites. Note that the sound waves in (b) have a much greater
amplitude (see the different density scales). }
\end{figure}

We first consider the behaviour of a dark soliton under harmonic
confinement only, in order to compare and contrast this to the
behaviour observed when an optical lattice is additionally
imposed. Figure 2(a) shows a space-time carpet plot of the
renormalised condensate density (actual density minus background
density) of a dark soliton, initially at the origin with speed
$v=0.5c$. The soliton oscillates back and forth in the trap at a
rate roughly equal to $\omega_s=\omega_x/\sqrt{2}$, as expected
\cite{fedichev,busch,huang,frantzeskakis}. As it climbs the side
of the trap, passing through the inhomogeneous background density,
it emits counter-propagating sound waves of opposing amplitude,
and becomes asymmetrically deformed \cite{parker2}. The sound
reflects off the edge of the condensate and subsequently
reinteracts with the soliton around $t\sim 5 \omega_x^{-1} $.  An
equilibrium is rapidly established where the soliton, on average,
reabsorbs as much sound as it emits.  Hence the soliton energy
features oscillations due to the emission/reabsorption process,
but with no net decay \cite{parker1}. The soliton position and
energy as a function of time are shown respectively by grey lines
in figure 3(a) and (b).

Consider now the case of imposing, in addition to the harmonic confinement,
 an optical lattice with height
$V_0=0.2\mu$ and wavelength $\lambda=2.4 l_x$, as shown in figure
2(b).  The presence of the optical lattice causes the soliton
dynamics to differ significantly from the pure harmonic case. As
the soliton oscillates, it not only climbs the walls of the
harmonic trap, but also traverses the lattice bumps. It thereby
experiences increased dynamical instability and emits additional
sound waves as it passes over each lattice site.  The harmonic
trap still contributes to the sound emission, as seen by the faint
`white' band of sound in figure 2(b), but this emission is heavily
obscured (compared to figure 2(a)) by the lattice-generated sound
waves. Importantly, in this case, reflection of sound from the
edge of the condensate no longer leads to the establishment of an
equilibrium between sound emission and reabsortion. This is
because the presence of the optical lattice dephases, or mixes,
the emitted sound such that only a proportion is reabsorbed. The
soliton becomes faster and shallower, and `anti-damps' to larger
amplitudes. By $t \sim 20 \omega_x^{-1}$ the soliton has becomes
so fast and shallow that it is virtually indistinguishable from,
and gradually merges into, the sound field. The soliton position
and energy as a function of time are shown respectively by dotted
lines in figure 3(a) and (b).

\section{Effect of varying the lattice height}

We now consider the effect on the soliton dynamics of varying the
height of the optical lattice $V_0$. Throughout this section, the
periodicity of the lattice is kept constant at $\lambda/2=1.2
l_x$. Figure 3 shows the soliton position and energy as a function
of time for various lattice heights. The soliton energy $E_s$ is
calculated by integrating the GP energy functional,
\begin{eqnarray}
\varepsilon(\psi)=\frac{\hbar^2}{2m}|\nabla
\psi|^2+V|\psi|^2+\frac{g}{2}|\psi|^4,
\end{eqnarray}
across the `soliton region', conveniently defined to be
$(x_s\pm5\xi)$, and subtracting from this the corresponding
contribution of the background fluid (obtained from the
time-independent solution of equation (1) in the absence of a
soliton).
\begin{figure}
\begin{center}
\includegraphics[width=9.5cm]{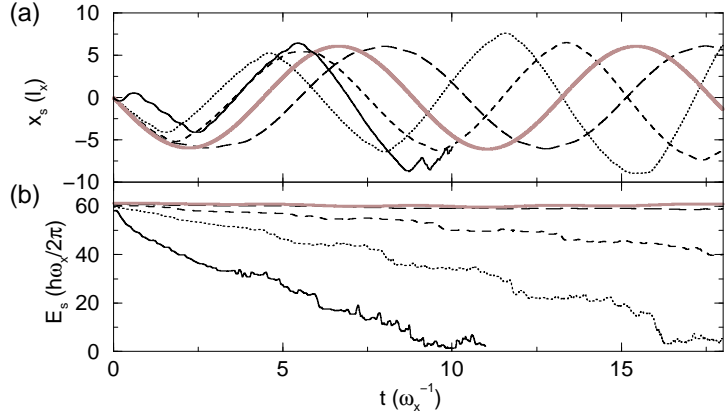}
\end{center}
\caption{Effect of lattice height on the soliton dynamics. (a)
Path and (b) energy of the soliton in a harmonic trap (solid grey
lines), and a harmonic trap plus an optical lattice with
$\lambda=2.4 l_x$ and lattice heights $V_0=0.05\mu$ (long-dashed
line), $V_0=0.1\mu$ (dashed line), $V_0=0.2\mu$ (dotted line) and
$V_0=0.4\mu$ (solid line). }
\end{figure}

Figure 3 shows soliton oscillations and energies for various
lattice heights.  Increasing the lattice height leads to two
distinct effects which modify the soliton oscillations in opposing
manners. Firstly, the optical lattice modifies the condensate
profile, periodically reducing the density. This inhibits the
motion of the soliton and tends to shift its turning point towards
the centre of the trap and {\it increase} its oscillation
frequency. This effect can be seen by the dashed line
($V_0=0.1\mu$) in figure 3(a). Indeed, if the lattice height is
sufficiently large, the soliton will initially be trapped within a
single lattice site. Secondly, the optical lattice induces sound
emission from the soliton.  The soliton energy decays essentially
monotonically, with the average decay rate increasing with lattice
height.  This decay leads to a gradual {\it decrease} in the
oscillation frequency and {\it increase} in the amplitude.  For
small lattice heights, e.g. $V_0=0.05\mu$ (long-dashed line in
figure 3(a) and (b)) the optical lattice poses little resistance
to the soliton motion, and the low level of sound emission it
induces causes the main observable effects. These are a small
decrease in the soliton oscillation frequency and a slow decay of
the soliton energy. The competing effects are most clearly
observed for $V_0=0.4\mu$ (solid line in figure 3(a)). The soliton
is initially confined to a single lattice site, with a high
oscillation frequency.  Sound emission causes the soliton to
gradually `anti-damp' out of this site into adjacent sites,
leading to a decrease in its oscillation frequency and an increase
in its amplitude. This effect has already been reported in
\cite{theocharis}. The soliton rapidly accelerates towards the
edge of the cloud where it ultimately merges into the sound field.
However, in the limit $V_0
> \mu$, an array of mini-condensates is formed.  Each lattice site
becomes an effectively closed system, and the emitted sound cannot
escape the site. In this limit the GP equation predicts that full
sound reabsorption by the soliton occurs, which stabilises it
against decay, as already observed in \cite{parker1}.  In the
limit of very small lattice periodicities, the discrete nonlinear
Schrodinger equation may be a more appropriate description, in
which case additional oscillatory instabilities come into play
\cite{johansson,kevrekidis}.

\section{Effect of varying the lattice periodicity}
\begin{figure}
\begin{center}
\includegraphics[width=8.0cm]{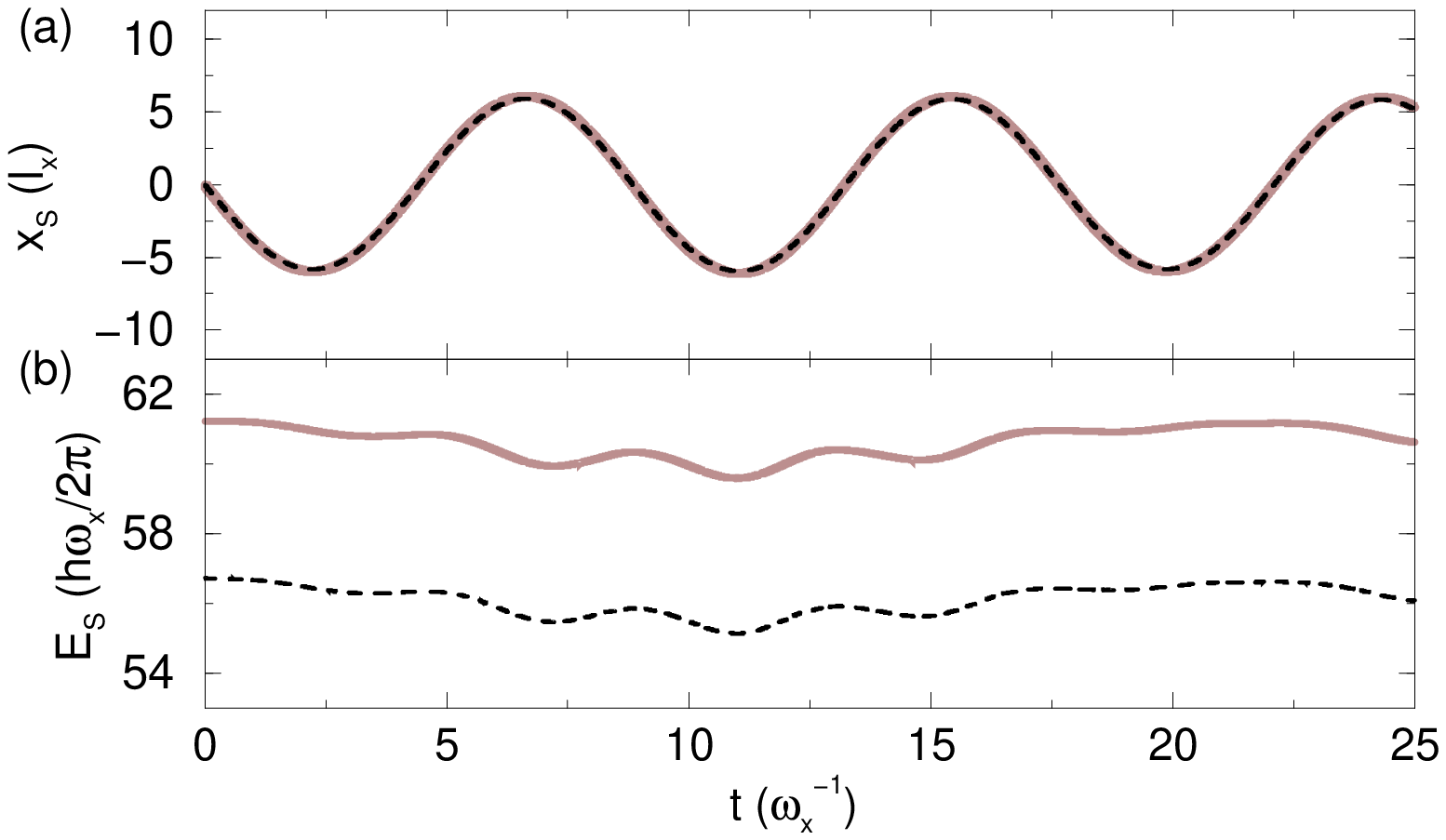}
\includegraphics[width=3.9cm]{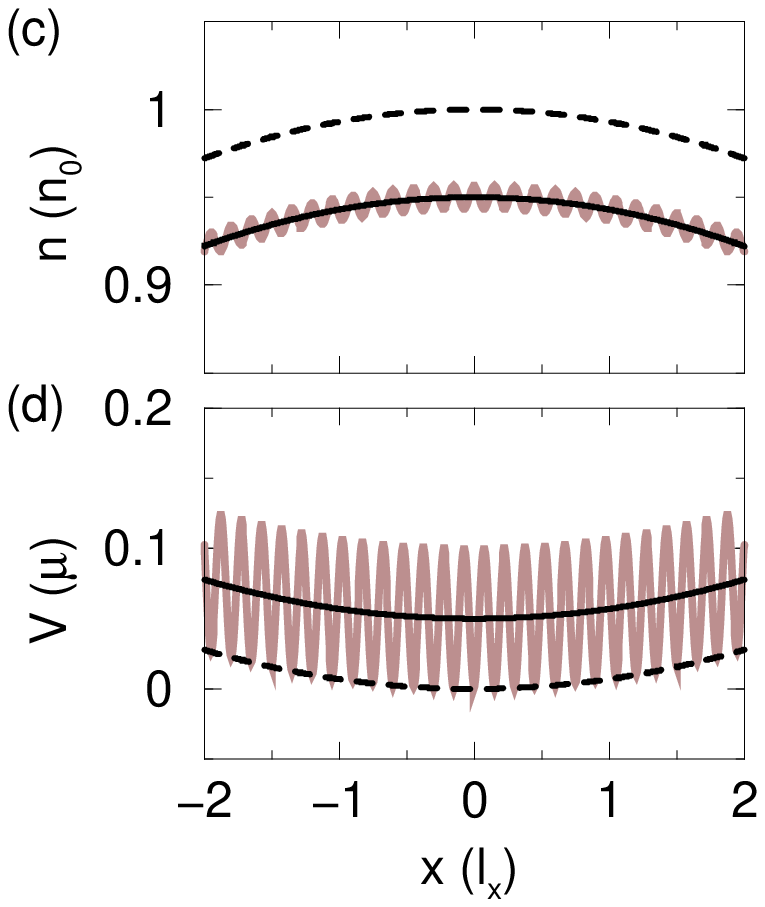}
\end{center}
\caption{Effect of small lattice periodicity, $\lambda/2 \sim \xi$
(or less), compared to no optical lattice. (a) Path and (b) energy
of a dark soliton in a tight optical lattice, $\lambda=0.3
l_x~(2.5 \xi)$, with height $V_0=0.1\mu$ (dashed line), compared
to the absence of the optical lattice (grey line). (c) The density
(grey line) for this tight lattice cannot fully heal, shifting the
average density down (bold line) in relation to the harmonic trap
density (dashed line). (d) Since the fluid cannot heal fully to
the tight lattice (grey line), the effective potential tends
towards a harmonic trap shifted upwards (solid line) by the mean
lattice potential $V_0/2$.  The harmonic potential is also shown
(dashed line).}
\end{figure}

\begin{figure}
\begin{center}
\includegraphics[width=6.3cm]{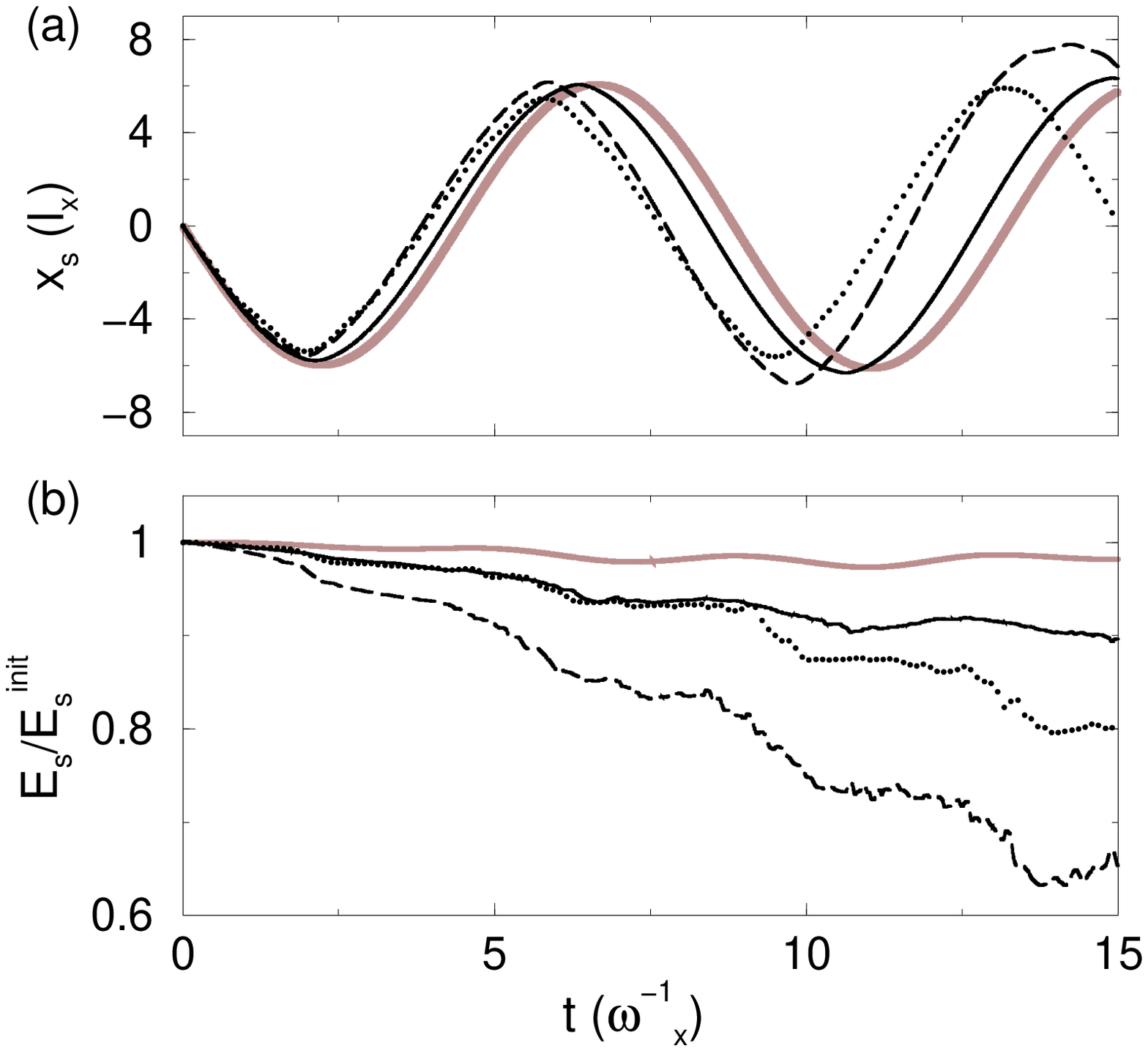}
\includegraphics[width=6.05cm]{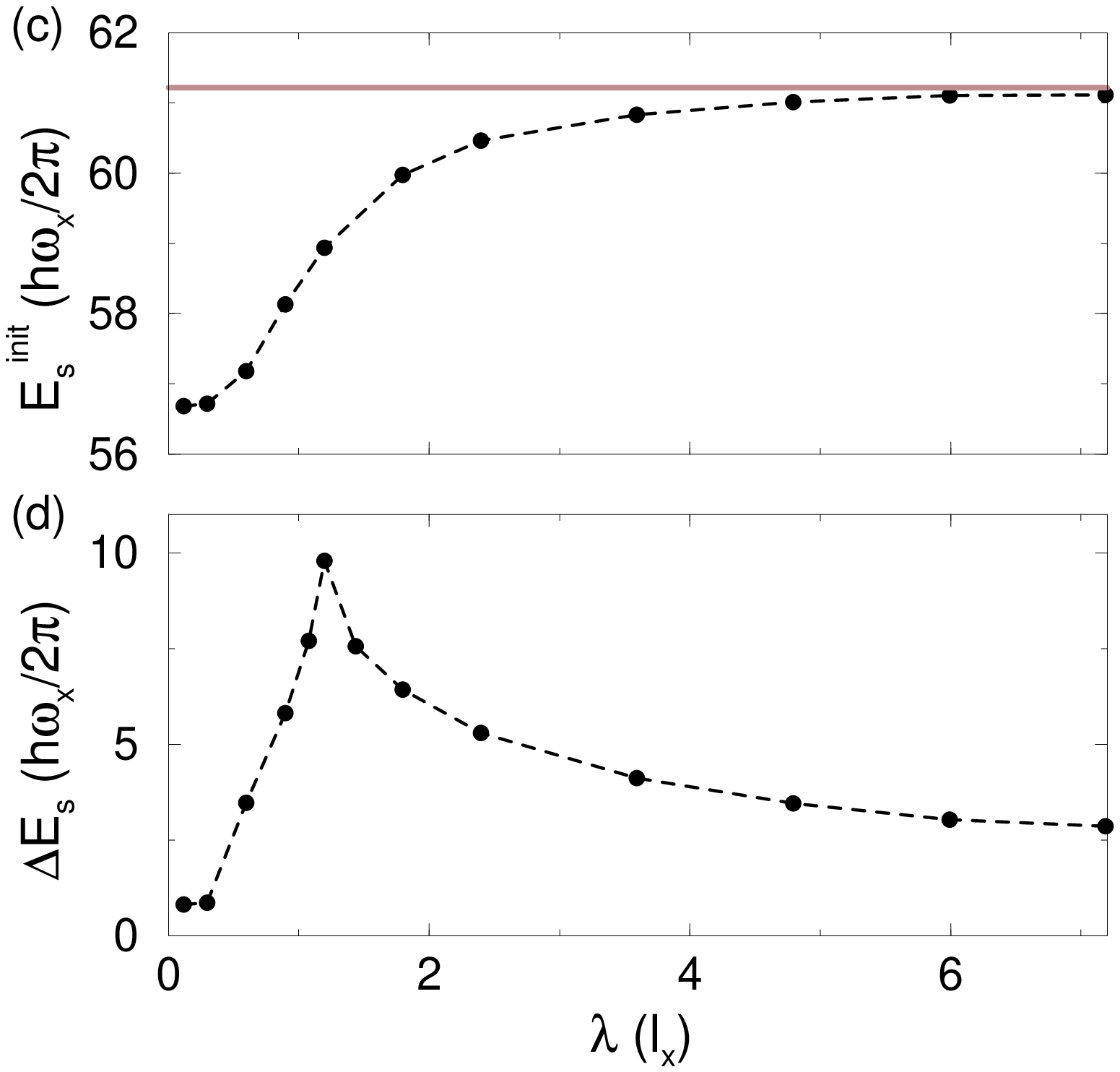}
\end{center}
\caption{Effect of lattice periodicity on the soliton dynamics.
(a) Path and (b) energy, rescaled by the initial soliton energy
$E_s^{\rm init}$, of a soliton in a harmonic trap (grey line) and
an additional optical lattice with height $V_0=0.1\mu$ and various
wavelengths: $\lambda=0.6 l_x~(5\xi)$ (solid line), $\lambda=1.2
l_x~(10\xi)$ (long-dashed), and $\lambda=4.8 l_x ~(40\xi)$ (dotted
line). (c) Initial soliton energy $E_s^{\rm init}$ as a function
of lattice wavelength, compared to the $V_0=0$ harmonic case
(horizontal grey line). (d) Loss of soliton energy $\Delta E_s$
after one full oscillation, as a function of lattice wavelength.}
\end{figure}

We next consider how the periodicity of the lattice affects the
soliton dynamics, an effect also considered analytically and
numerically in \cite{theocharis}.  We start by discussing an
anomalous regime, the tight lattice regime, where the lattice
periodicity is of the order of, or less than, the soliton width.
The path and energy of a dark soliton with initial speed $v=0.5c$
under the influence of a tight optical lattice ($\lambda=0.3 l_x
~(2.5\xi)$
 and $V_0=0.1\mu$) are shown by the dashed line in figure 4(a)
and 4(b) respectively. The path of the soliton is identical to
that in the absence of the optical lattice (solid grey line),
while the energy has the same form but is shifted to a lower
value. This effect can be explained by careful consideration of
the density perturbations caused by the optical lattice. In the
limit where the lattice wavelength is of the order of, or less
than, the soliton width, the fluid cannot heal to the optical
lattice, and the associated density modulations smear together.
Then the background density becomes reduced across the whole
condensate, and in particular the peak density, $n_0\rightarrow
n'_0$. This is due to the fact that, in our simulations, the
chemical potential $\mu$ is fixed. If, however, the peak
density/number of particles were kept fixed, then the chemical
potential would increase accordingly. The soliton essentially sees
a constant effective optical lattice potential of height $V_0/2$,
and overall it sees the ambient harmonic potential shifted upwards
by this amount, an effect illustrated in figure 4(c) and 4(d). In
both cases of a pure harmonic trap and a harmonic trap modified by
a tight optical lattice, the initial soliton speed has been set to
the same value $v=0.5c$ (i.e. the dependence on peak density is
removed), ensuring that the paths are identical. If we assume a
Thomas-Fermi relationship between potential and density,
$n(x)=n_0-V(x)$, then we would expect, to first order, the peak
density to be decreased by $(V_0/2)n_0=0.05n_0$. Equation 4
implies that the initial soliton energy will be rescaled by a
factor of $(n'_0/n_0)^\frac{3}{2}$, i.e. $E_s^{\rm init}
\rightarrow 0.926 E_s^{\rm init}$, and this is indeed the modified
soliton energy observed.

We now investigate the effect of lattice periodicity on the
soliton dynamics. The soliton path and energy for various lattice
wavelengths are shown in figure 5(a) and (b) respectively. In
order  to remove the effect of reduced peak density for
sufficiently tight lattices (based on our fixed chemical potential
simulations), the soliton energies in figure 5(b) have been
rescaled in terms of the initial soliton energy.  This effect is
quantified in figure 5(c) which shows the variation in the initial
soliton energy as a function of lattice wavelength.

Looking now at the decay of the (normalised) soliton energy as a function of time
for various lattice wavelengths, we find that increasing the
 lattice wavelength (but still remaining far from the tight lattice
regime considered earlier) leads to the decay of the soliton
energy and an associated increase in the soliton amplitude and
frequency (e.g. $\lambda=0.6 l_x$, solid line in figure 5(a)
and(b)) in relation to the pure harmonic case (grey line). The
rate of this decay increases up to some resonant wavelength
($\lambda=1.2 l_x$, dashed line). Any further increase in the
wavelength leads to a slowing of the decay.  This is illustrated
in figure 5(b) by the reduced decay rate for $\lambda=4.8\xi$
(dotted line) compared to $\lambda=1.2 \xi$ (dashed line).

The dependence of the soliton decay on the lattice periodicity can
be explained in terms of the density inhomogeneity that the
soliton experiences. To illustrate this analysis more
quantitatively, figure 5(d) plots the loss in soliton energy after
one full oscillation as a function of lattice wavelength. For
tight lattices, where the lattice periodicity is of the order of,
or less than, the soliton width, the soliton effectively sees a
displaced harmonic trap, and so experiences no net decay, due to
complete sound reabsorption. In this limit, the soliton energy
oscillates periodically due to the continuous sound
emission/reabsorption process.  This takes the form a beating
effect caused by the different oscillation frequencies of the
soliton and sound (dipole mode) \cite{parker1}. Therefore, one
full oscillation of the soliton in the trap does not correspond to
one beat period, and hence the soliton energy has not yet returned
to its initial value, explaining the small apparent energy loss.
For progressively weaker lattices, the density begins to heal to
the optical lattice, thereby modifying the condensate density from
the harmonic trap profile. This increases the density
inhomogeneity that the soliton sees, and induces an increase in
the rate of decay. Furthermore, the presence of these density
modulations dephases the sound field such that the soliton is no
longer fully stabilised by sound reabsorption.

Increasing the lattice periodicity further leads to some critical
wavelength the condensate density can heal fully to the lattice
potential. This represents the system with the maximum density
inhomogeneity (for a fixed lattice height) and therefore features
the most rapid soliton decay. This point is indicated by the
resonance in figure 5(d) and occurs for $\lambda \sim 1.2
l_x~(10\xi)$. Further increase in the wavelength does not change
the depth of the density modulations, but merely stretches them
out. This now reduces the density gradient experienced by the
soliton \cite{parker2} and causes a reduction in the decay rate.
As $\lambda \rightarrow \infty$ the decay will asymptotically tend
towards zero as the trap potential returns to becoming effectively
harmonic.

\section{Quantifying the sound emission}
We will now quantatively discuss the soliton dynamics up until the
time when the sound reinteracts with the soliton. Figure 6 shows
the short-term dynamics of a dark soliton in the optical lattice
geometry considered in Section 3 ($V_0=0.2\mu, \lambda=2.4 l_x$,
dotted line of figure 3), along with the corresponding dynamics in
the absence of the optical lattice (grey line of figure 3). The
soliton motion (figure 6(a)) is clearly inhibited by the modified
local density (figure 6(b)), causing it to turn around earlier and
at a lower amplitude than in the absence of the optical lattice.
The soliton accelerates and decelerates periodically as it
traverses the lattice sites, as shown in figure 6(c) (solid line,
left axis) with the oscillation weakly modulated by the additional
acceleration induced by the ambient harmonic trap (grey line in
same figure). This contribution is zero when the soliton is at the
centre of the trap and a maximum when the soliton is at an
extremum.

The emission of oppositely ``charged'' sound waves from the
soliton leads to an asymmetric distortion of the soliton
profile, and a shift of the soliton centre of mass $x_{\rm cm}$
from the density minimum $x_s$, where the soliton centre of mass
is defined as,
\begin{eqnarray}
x_{\rm cm}=\frac{\int_s x(|\psi|^2-n) {\rm d}x}{\int_s
(|\psi|^2-n) {\rm d}x}.
\end{eqnarray}
This shift, shown in figure 6(c) (dashed line, right axis), is
directly proportional to the soliton acceleration (solid line,
left axis), in accordance with earlier findings \cite{parker2}.

\begin{figure}
\begin{center}
\includegraphics[width=13cm]{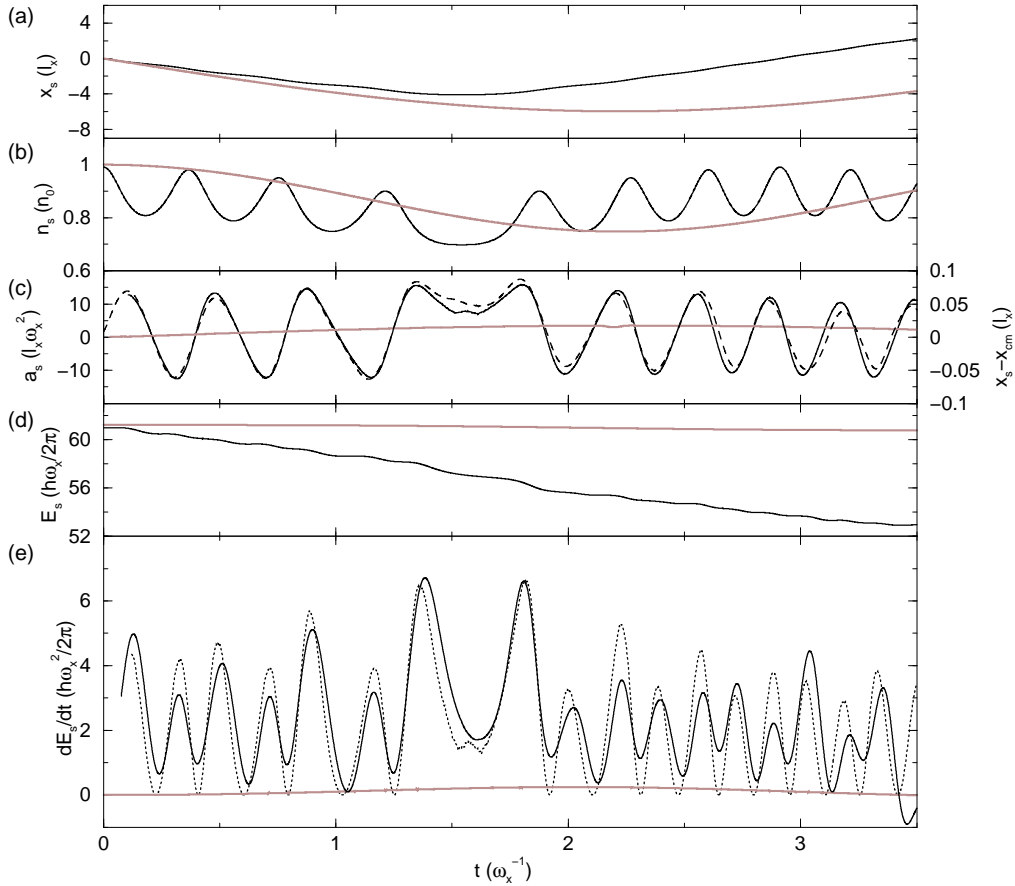}
\end{center}
\caption{Comparison of dynamics of a soliton with initial speed
$0.5c$ in a harmonic trap plus optical lattice with $V_0=0.2\mu$
and $\lambda=20\xi$ (black line) and corresponding dynamics only
in the presence of the same harmonic trap (grey lines). (a) Path
of soliton versus time. (b) Background density at the soliton
position. (c) Soliton acceleration (bold line, left axis) and
deformation parameter $(x_s-x_{\rm cm})$ (dashed line, right
axis). (d) Soliton energy. (e) Power emitted by the soliton as
computed from the GP energy functional (solid line), compared with
the acceleration squared law of equation (7) with $\kappa=1.8
(\hbar/l_x)$ (dotted line).}
\end{figure}

The motion of the soliton through the lattice causes a periodic
cascade of the soliton energy (figure 6(d)).  This cascade is more
evident when considering the power emitted by the soliton, shown
in figure 6(e).  The power consists of a series of peaks of
various amplitudes, which clearly occur whenever the soliton
traverses the side of a lattice site, i.e. where the density
gradient is a local maximum. The peaks are greatest at around $t
\sim 1.5 \omega_x^{-1}$, which corresponds to the point of maximum
soliton acceleration. Indeed, the power emission can be well
fitted by an acceleration-squared law,
\begin{eqnarray}
\frac{dE}{dt}=-\kappa a^2
\end{eqnarray}
where $\kappa$ is a constant coefficient ($\kappa=0.0255
(\hbar/l_x)$ in figure 6(e)). This is also in agreement with an
acceleration-squared power law derived analytically for
dynamically unstable dark solitons in the context of nonlinear
optics \cite{pelinovsky}. Although \cite{pelinovsky} deals with a
homogeneous system in which the instabilities arise from
imperfections in the medium, such an approach has recently been
considered for dark solitons in inhomogeneous condensates
\cite{parker1,parker2}. When the emitted sound reinteracts with
the soliton ($t > 3 \omega^{-1}_x$), it is partially reabsorbed
(as evident from the negative values acquired by the power curve
of figure 6(e)), and equation (7) ceases to be valid.  For the
same reasons, the proportionality between acceleration and the
deformation parameter is no longer valid.

\section{Conclusions}

We have made a quantitative investigation of the dynamics of a
dark soliton induced by the presence of an optical lattice. We
find, in general, that the optical lattice significantly affects
the motion and dynamical stability of the soliton. The exception
to this is in the tight lattice regime where the fluid cannot heal
to the lattice and the soliton effectively sees a harmonically
trapped condensate.

By considering optical lattices of varying depths and wavelengths
we have illustrated the sensitive dependence of the soliton
dynamics on these parameters, and explained the dependence in
terms of the density inhomogeneity experienced by the soliton. The
effect of the optical lattice, and the associated density
modulations, is to create extra dynamical instability in the
soliton, in addition to that caused by the ambient harmonic trap.
This induces soliton decay via the emission of sound waves.  For
deep and moderately tight lattices this effect is considerable:
the soliton rapidly decays, accelerating to the edge of the
condensate, where it disappears into the sound field.  Unlike in a
pure harmonic trap, the presence of the optical lattice tends to
modify the emitted sound such that the full stabilisation of the
soliton by sound reabsorption is prevented.

The soliton-sound interaction in an optical lattice geometry can
be experimentally probed in a quasi-one-dimensional condensate,
providing an alternative method to that suggested in
\cite{parker1}. The dynamical dissipation is observable as a
change in the soliton amplitude and oscillation frequency, by
repeated expansion of the cloud and subsequent absorption imaging,
after a variable evolution time in the trap.  The timescales
considered in this work are consistent with the expected
thermodynamic soliton lifetime of 1s \cite{fedichev}.  The
intensity of the laser beams that form the optical lattice
controls the lattice height, while varying the relative angle
between the beams allows the full range of lattice periodicities
considered here to be realised.  Modification of these two
parameters enables full experimental control of the dynamical
instability of the soliton induced by the optical lattice.

\ack We acknowledge financial support from the UK EPSRC and
discussions with D. J. Frantzeskakis and P. G. Kevrekidis.

\end{document}